# Multi-Personality Partitioning for Heterogeneous Systems

Anthony Gregerson, Aman Chadha, Katherine Morrow
Department of Electrical & Computer Engineering
The University of Wisconsin – Madison
Madison, WI USA
agregerson@wisc.edu, achadha2@wisc.edu, kati@engr.wisc.edu

*Abstract*—Design flows use graph partitioning both as a precursor to place and route for single devices, and to divide netlists or task graphs among multiple devices. Partitioners have accommodated FPGA heterogeneity via multi-resource constraints, but have not yet exploited the corresponding ability to implement some computations in multiple ways (e.g., LUTs vs. DSP blocks), which could enable a superior solution. This paper introduces multi-personality graph partitioning, which incorporates aspects of resource mapping into partitioning. We present a modified multi-level KLFM partitioning algorithm that also performs heterogeneous resource mapping for nodes with multiple potential implementations (multiple personalities). We evaluate several variants of our multi-personality FPGA circuit partitioner using 21 circuits and benchmark graphs, and show that dynamic resource mapping improves cut size on average by 27% over static mapping for these circuits. We further show that it improves deviation from target resource utilizations by 50% over post-partitioning resource mapping.

## I. Introduction

Netlist partitioning is a critical part of the CAD flow for large circuit designs, both to achieve high-quality results and to reduce the CAD flow execution time. Existing research on multi-resource partitioning for heterogeneous FPGAs has focused on partitioning nodes with pre-assigned resource costs [1] [2] [3]. However, heterogeneity presents the opportunity to implement some computations in multiple different ways, using different combinations of resources. For example, multiply-accumulate operations can be implemented in DSP blocks or CLBs. We refer to logic with multiple possible implementations as ***multi-personality*** logic, and the process of choosing an implementation for that logic as ***personality selection***. CAD tools perform personality selection during synthesis; however, allowing a post-synthesis partitioner to modify personality selections can improve partitioning both in terms of cut size and heterogeneous resource utilization.

In this paper we define the multi-personality partitioning problem and propose specific modifications to multi-level KLFM (Kernighan-Lin/Fiduccia-Mattheyses) to incorporate personality selection. We evaluate our modifications using a collection of graphs based on real-world multi-FPGA designs and publicly-available benchmarks. We demonstrate that they result in better cut sizes and deviations from target resource utilizations than partitioning of fixed-implementation netlists.

## II. Related Work

Circuit netlists can be represented as hypergraphs, where logic elements (or sometimes groups of elements) are graph nodes and the nets that connect the nodes are hyperedges. Netlist/Hypergraph partitioning based on variants of the KLFM algorithm [4] often optimize for metrics such as channel width/congestion, ease of routability, and operating frequency [5]. Some work has also examined multi-resource constraints for heterogeneous devices for KLFM [2] and other algorithms [3], but unlike our work, they do not integrate personality selection for multi-personality nodes into partitioning. In general, our proposed changes do not conflict with other KLFM partitioning extensions. Furthermore, some of our techniques, such as phase-based implementation rebalancing and multi-level implementation control, could also be adapted to non-KLFM-based partitioning algorithms.

## III. Problem Definition

We define graph partitioning for heterogeneous devices as follows: Partition a graph with nodes that have weights in $R$ resources such that all $R$ resources are balanced within their specified maximum imbalance margins. Each node's weight is now described by an $R$-entry weight vector $w$. If we define $w_r(v)$ as node $v$'s weight in resource $r$, and $I_r$ as the maximum weight imbalance between partitions for resource $r$, then

$$\left| \sum_{v_i \in V_1} w_r(v_i) - \sum_{v_j \in V_2} w_r(v_j) \right| \leq I_r \quad \forall \ r = 1..R \quad (1)$$

The resulting cut sizes for multi-resource partitioning are usually higher than those of single-resource partitioning due to the additional $R-1$ constraints.

For devices like FPGAs a given node may have multiple implementations using different resource weights. We thus define a new *multi-personality graph partitioning* problem. If we define all possible implementations of node $v$ as $P_v$ and the selected personality as $p(v)$, we can reformulate the partitioning constraints as:

$$\left| \sum_{v_i \in V_1} w_r(p(v_i)) - \sum_{v_j \in V_2} w_r(p(v_j)) \right| \leq I_r \quad \forall \ r = 1..R \quad (2)$$

Whereas multi-resource partitioning has $O(2^N)$ potential solutions, where $N$ is the number of nodes, multi-personality partitioning has $O(C(P) \times 2^N)$ potential solutions, where

$$C(P) = \prod_{v \in V} |P_v| \qquad (3)$$

The number of personality combinations scales non-polynomially with the number of multi-personality nodes, resulting in a much larger solution space. Although this flexibility may make it possible to find superior solutions, it also increases the problem complexity. Furthermore, within each partition, we may also want to achieve a certain relative utilization of resources, for example to find the smallest usable FPGA in a given family to implement the netlist's partitions. We refer to this as the *resource utilization ratio* (RUR), and make it an additional goal for multi-personality partitioning.

## IV. MULTI-PERSONALITY PARTITIONING ALGORITHM

The base KLFM algorithm starts with an initial partitioning solution, and moves nodes between partitions in multiple passes. Each pass, nodes swap between partitions one-by-one until all that can move have moved, and at the end of each pass, the best solution achieved during that pass is kept as the starting point of the next pass. In single-resource KLFM, structures called *gain buckets* are used to find the highest-gain node (i.e., one that most reduces cutsize or other cost measures) that can move without violating balance constraints. The partitioning quality often decreases for very large graphs, however, so *multi-level KLFM* [6] hierarchically clusters graph nodes, and partitions at each level (starting with the coarsest) before un-clustering at that level and partitioning the next.

This section describes our modifications to the multi-level KLFM partitioning algorithm to support multi-personality graph partitioning. The required changes affect key data structures and introduce unique complexities. As part of our modifications, we added dynamic personality selection, experimented with modifications to the gain buckets, and introduced new pass-level global remapping. We also examined these techniques' effects on multi-level partitioning.

### A. Dynamic Gain Buckets

Multi-personality gain buckets also perform personality selection, with the goal of meeting balance and RUR goals. We experimented with two different approaches, described here.

*1) Multi-Personality Buckets*: Multi-Personality Buckets are similar to standard gain buckets, but each node's entry has a separate set of weights (indicating its resource requirements) for each of its personalities. Each step within a pass, we use a tournament-style selection policy to determine which node to move, selecting the highest-gain node with an implementation that would allow the node to move without violating constraints. For all of that node's implementations that do not violate constraints, we compute an *imbalance score*, defined as the root-mean-square (RMS) of the resulting fractional imbalances in all resources. The RMS tends to emphasize large imbalances in one resource over small imbalances in many resources; large imbalances are more likely to prevent future high-gain moves. To incorporate an RUR goal, we compute a similar *resource-ratio score* for each possible implementation as the RMS of the percent deviation of each resource from its target utilization, based on total utilization of all resources in a partition and the target utilization ratio. The node with the best imbalance and/or resource ratio score is selected for movement (depending on the partitioner policy).

*2) Resource-Affinity Buckets*: Sometimes it is beneficial to move nodes with implementations in specific resources. Resource-Affinity Buckets separate a partition's buckets into several resource-specific queues, making it easier to adjust a specific resource to improve RUR or alleviate partition imbalances. Resource-Affinity buckets can be used in place of Multi-Personality buckets, or combined in a hybrid approach at the cost of additional computational overhead.

### B. Pass-Level Implementation Remapping

Global remapping during partitioning can modify personality selections to reduce resource imbalance and/or improve RURs. Because it is a relatively expensive operation we apply $O(N)$-complexity global remapping algorithms once per pass, and thus do not increase the asymptotic complexity of the base KLFM algorithm. We experimented with two global remapping methods: a multi-phase greedy algorithm that performs a random walk on all nodes and changes their implementations to minimize imbalance and RUR scores, and a fractured integer linear program (ILP) method.

### C. Multi-Level Partitioning

Implementation selection is a major problem in multi-level KLFM, as each additional component node creates a geometric increase in the number of implementations available to a supernode. We use a similar approach to that described for populating Resource Affinity Buckets to determine the subset of implementations for each supernode; these implementations provide the partitioner with the greatest ability to adjust resource balance. We also select implementations with RURs closest to the target ratio, since achieving a similar weight for all nodes can at times lead to better partitioning results [7].

During experimentation, we observed that the partitioner rarely changed supernode implementations at the coarsest levels of multi-level partitioning, and that enforcing multi-resource constraints at these levels often led to poor results both at these and later levels. To address this, we relax resource constraints based on the coarseness of the partitioning level to increase freedom and improve results at the coarsest levels. As the algorithm progresses to less coarse graphs, stricter balance constraints are re-enforced.

### D. Multi-Personality Partitioning Strategies

We developed several different approaches for partitioning multi-personality, heterogeneous-resource graphs using the techniques that we have described earlier in this paper. When choosing these partitioning strategies, our goal was to select those that were representative of the options currently available for partitioning and implementation mapping algorithms. We also aim to highlight tradeoffs in multi-personality partitioning.

*1) Statically-Mapped Partitioning (SM):*
This strategy is based on the Native Multi-Constraint Refinement partitioning method [2] proposed for heterogeneous FPGAs, modified to use Resource-Affinity Buckets. We first apply our global implementation remapping algorithms to achieve the target RUR. Implementations are then fixed, and the graph is partitioned using multi-constraint KLFM without dynamic personality selection or remapping.

*2) Statically-Partitioned Remapping (SP):*
Statically-Partitioned Remapping converts as many nodes as possible into the commonly-usable resource (CLBs in the case of FPGAs) before partitioning. After partitioning, it globally remaps each partition based on the target RUR and balance constraints. SP is conceptually similar to the Multi-Constraint Iterative k-way Balancing method [6], however, it exploits nodes' multiple personalities and our global remapping algorithms to enforce constraints.

*3) Dynamic Multi-Personality Partitioning (DMP):*
DMP uses move-based implementation selection using Resource-Affinity Buckets and pass-based global remapping using the multi-phase greedy algorithm. Implementation selection is based on imbalance score. It uses fractured ILP global remapping to improve the RUR after partitioning.

TABLE I: SUMMARY OF BENCHMARK CIRCUITS/GRAPHS.

| Benchmark | Total Nodes | % CLB Nodes | % DSP Nodes | % BRAM Nodes |
|---|---|---|---|---|
| 144 | 144649 | 94 | 20 | 1 |
| 598a | 110971 | 98 | 18 | 0.5 |
| blob | 11842 | 100 | 24 | 0 |
| boundtop | 29582 | 100 | 0.5 | 14 |
| brack2 | 62631 | 99 | 1 | 8 |
| cti | 16840 | 97 | 14 | 2 |
| diffeq1 | 4292 | 100 | 38 | 0 |
| fe_ocean | 143437 | 97 | 3 | 3 |
| fe_rotor | 99617 | 99 | 2.5 | 1.5 |
| fe_tooth | 78136 | 98 | 4 | 4 |
| fft128 | 91590 | 100 | 9.5 | 0.5 |
| isolation | 187766 | 100 | 1 | 0 |
| jet | 189579 | 100 | 32 | 0 |
| m14b | 214765 | 98 | 3 | 2 |
| mcml | 346248 | 100 | 15 | 0.1 |
| memplus | 17758 | 98 | 2 | 40 |
| raygen | 11457 | 100 | 25 | 0.1 |
| rct | 241349 | 100 | 31 | 12 |
| sha | 3669 | 100 | 15 | 0 |
| wave | 156317 | 98 | 6 | 5 |
| wing | 62032 | 99 | 4 | 2.5 |

*4) Advanced Dynamic Multi-Personality (ADMP):*
ADMP expands on DMP, using Hybrid Buckets for move-based selection and multi-level constraint relaxation.

*5) DMP/ADMP with Fine-Grained Ratio Control (DMP-FR / ADMP-FR):*
This modification considers RUR score during move-based personality selection and pass-based global remapping, improving resource utilization at the cost of cut size.

V. EXPERIMENTAL METHODOLOGY

We implemented a complex-constraint, multi-level KLFM partitioner and modified it to include the multi-personality-aware features described in Section IV. Our base algorithm includes common KLFM optimizations such as LIFO gain bucket queues and randomized initial partitions [4]. We validated performance of the base algorithm using results from the Walshaw partitioning archive [7].

We then compared the results of our multi-personality algorithm against non-personality-aware partitioning approaches in both cut size and RUR using a suite of 21 benchmark graphs, summarized in Table I. These include several heterogeneous netlists for FPGAs from ERCBench [8] and VPR [9]; graphs from the Walshaw graph partitioning archive from domains such as communications, vibroacoustics, and 3D meshes [7]; several FPGA circuits designed for multi-FPGA high-energy physics experiments such as particle isolation (isolation), jet reconstruction (jet), and calorimetry-based triggering (rct) [10]; a ray tracing circuit (raygen) [11]; and a circuit for Monte Carlo simulation for photodynamic cancer therapy (mcml) [12]. Where HDL was available, we assigned node weights and personalities using experimental synthesis in Xilinx ISE and fixed implementations of nodes along critical paths, since these may not have the flexibility to use alternate personalities due to timing constraints. For benchmarks distributed as graphs, we computed a range of potential resource costs for functional units by synthesizing a range of DSP and BRAM-based units and assigned weights within that range using a binomial distribution.

For each experiment, we report the best result from 50 runs, since KLFM partitioners often use the best result of multiple runs. We applied an imbalance limit of 1% for all resources. We set a target ratio of 1:1:1 for the percent utilization of the three primary FPGA resources and computed the root-mean-square of the percent deviation from the target resource weight. If a circuit did not use a resource, it was ignored. Resource capacity ratios are based on the Xilinx Virtex-7 2000T [13].

VI. RESULTS

Fig. 1 shows a comparison of cut size results for 2-way multi-personality partitioning, normalized to the result of Statically-Mapped Partitioning (SM). Cut size is reported as the sum of the weight of all wires that span more than one partition. Due to space limitations, the chart includes a representative set of individual results for nine benchmarks, and the geometric mean of results for all 21 benchmarks. Fig. 2 shows a comparison of target resource utilization deviation. The results are also summarized in Table II. For each metric shown in these figures and this table, lower values are better.

*A. Cut Size and Target Resource Utilization*

SM achieves worse cut sizes than other strategies due to its difficult task of balancing multiple resources. Results of the dynamic algorithms demonstrate that changing node personalities during partitioning can improve cut sizes, resulting in mean improvements of 27% over SM and 6% over SP when using ADMP. The addition of dynamic ratio control harmed cut size results by a small amount, but improved deviations from the target RUR. SM, however, results in the best deviation from target RUR because its netlists are mapped prior to partitioning and each resource is limited to a maximum imbalance of 1% between partitions. However, this comes at the aforementioned cut size penalty. SP had the worst deviation—not using personality information during portioning makes it difficult to balance heterogeneous resource utilization afterwards—which in almost all cases came from over-utilizing CLBs and under-utilizing specialized resources. Results demonstrate that the dynamic algorithms offer good cut sizes and a compromise in deviation between SP and SM.

For both cut size and RUR deviation, the magnitude of the difference between the partitioning strategies varied heavily from benchmark to benchmark, largely as a consequence of the topology and the quantity and distribution of resources in the graph. The circuit for particle isolation, for example, has a very regular grid-based structure with uniform resource distribution, so almost any partitioning would lead to balanced resource utilization. Jet and boundtop, on the other hand, had most of their non-CLB resources distributed in clusters, allowing for more significant tradeoffs between balanced resource utilization and cut size. The multi-level constraint relaxation technique included in DMP+ proved particularly effective with the highly clustered and interconnected topologies of large memories, as evidenced by its improvement in cut size for boundtop, which includes a large shared memory.

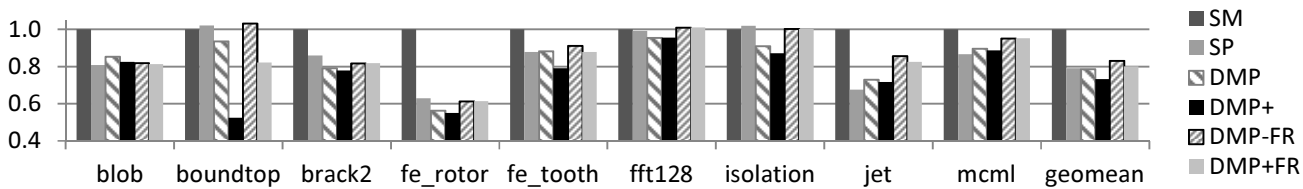

Fig. 1. Cut size results, normalized to statically-mapped partitioning. The geometric mean includes data from all 21 benchmarks. Lower values are better. These results are also summarized in Table II.

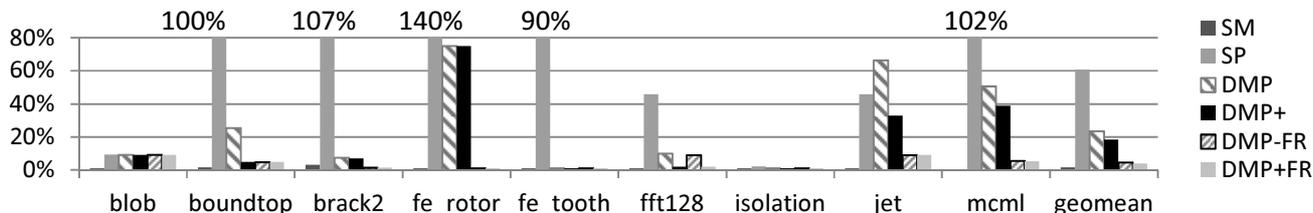

Fig. 2. RMS deviation from target resource utilization for a subset of benchmarks. The geometric mean contains data from all 21 benchmarks. Lower values are better. These results are also summarized in Table II.

TABLE II: GEOMETRIC MEANS OF THE EVALUATED COST METRICS AND NORMALIZED WALL-CLOCK RUN TIME FOR EACH STRATEGY.

|  | SM | SP | DMP | ADMP | DMP-FR | ADMP-FR |
|---|---|---|---|---|---|---|
| Norm. Cut Size | 1.00 | 0.79 | 0.78 | 0.73 | 0.83 | 0.80 |
| RUR Deviation | 1.6% | 61% | 24% | 19% | 4.6% | 4.0% |
| Norm. Run Time | 1.0 | 0.6 | 1.3 | 1.7 | 1.3 | 1.7 |

### B. Run-Time Cost

Although our modifications do not alter the asymptotic complexity of the base multi-level KLFM algorithm, they increase algorithm run-time linearly. To roughly estimate the impact of various techniques on run time, we ran the multi-personality algorithm with different combinations of enabled features over a subset of the largest benchmarks and recorded the wall clock time. Table II presents the results, normalized to the run time of Statically-Mapped Partitioning.

## VII. CONCLUSION

Although existing partitioning algorithms can handle multiple resources on heterogeneous devices in a primitive fashion, they do not exploit the overlapping functionality provided by some of these resources. Integrating dynamic selection of node implementations into the partitioner achieves up to a 27% mean improvement in partition cut size compared to partitioning a statically-mapped circuit for our suite of 21 benchmarks. Dynamically selecting node personalities for ratio control can achieve up to a 15X mean improvement in deviation in target resource utilization compared to post-partition resource mapping. Combining the enhanced dynamic algorithm with fine-grained resource ratio control makes it possible to achieve better cut size benefits than statically partitioned enforcement-based strategies while also achieving most of the resource utilization advantages of statically-mapped partitioning approaches.